\documentclass[preprint,showpacs,preprintnumbers,amsmath,amssymb]{revtex4}

\usepackage{graphicx}
\usepackage{dcolumn}
\usepackage{bm}
\usepackage{amsmath,dsfont}
\usepackage{amssymb,amsthm,amscd, amsbsy, array}
\usepackage{graphics,graphicx,xcolor}
\usepackage{etex}
\usepackage[all]{xy}

\numberwithin{equation}{section}

\newcommand{\dsp}{\displaystyle}

\newcommand{\hs}[1]{\hspace{#1 em}}
\newcommand{\bflr}{\begin{flushright}}
\newcommand{\eflr}{\end{flushright}}
\newcommand{\bc}{\begin{center}}
\newcommand{\ec}{\end{center}}
\newcommand{\ben}{\begin{enumerate}}
\newcommand{\een}{\end{enumerate}}

\newcommand{\be}{\begin{equation}}
\newcommand{\ee}{\end{equation}}
\newcommand{\ba}{\begin{array}}
\newcommand{\ea}{\end{array}}
\newcommand{\ct}{\cite}

\newcommand{\dd}[2]{\frac{\partial{#1}}{\partial{#2}}}

\newcommand{\eps}{\epsilon}

\newcommand{\rg}{\rho}

\newcommand{\vf}{\varphi}
\newcommand{\og}{\omega}
\newcommand{\Gam}{\Gamma}

\newcommand{\Fg}{\Phi}

\newcommand{\bfpi}{\mbox{{\boldmath $\pi$}}}

\newcommand{\bfR}{\mathbf{R}}

\newcommand{\lh}{\left(}
\newcommand{\rh}{\right)}
\newcommand{\ld}{\left.}

\newcommand{\der}{\partial}

\newcommand{\half}{{\scriptstyle{\frac{1}{2}}}}
\def\2{{\half}}
\def\smallcirc{{\,\raise 0.5pt \hbox{$\scriptstyle\circ$}\,}}

\newcommand{\orth}{{\rm o}}

\def\beq{\begin{equation}}
\def\eeq{\end{equation}}
\def\beqa{\begin{eqnarray}}
\def\eeqa{\end{eqnarray}}

\def\barray{\left(\begin{array}}
\def\earray{\end{array}\right)}
\def\barraynb{\begin{array}}
\def\earraynb{\end{array}}
\def\benu{\begin{enumerate}}
\def\eenu{\end{enumerate}}
\def\bea{\begin{eqnarray}}
\def\eea{\end{eqnarray}}

\def\smallover#1/#2{\hbox{$\textstyle\frac{#1}{#2}$}} %

\def\benu{\begin{enumerate}}
\def\eenu{\end{enumerate}}

\usepackage{color}

\begin{document}

\preprint{arXiv:1401.8195v3
}

\title{Killing tensors and canonical geometry
}

\author{
M. Cariglia$^{1}$\footnote
{e-mail:marco@iceb.ufop.br}
G.~W.Gibbons$^{2}$\footnote{mail:G.W.Gibbons@damtp.cam.ac.uk},
J.-W. van Holten$^{3,4}$\footnote{
mail: t32@nikhef.nl},
P.~A.~Horvathy$^{5,6}$\footnote{
e-mail:horvathy-at-lmpt.univ-tours.fr},
P. Kosi\'nski$^{7}$\footnote{email: pkosinsk@uni.lodz.pl},
P.-M. Zhang$^{5,8}$\footnote{e-mail:zhpm@impcas.ac.cn}
}

\affiliation{
$^{1}$DEFIS, Universidade Federal de Ouro Preto, Campus Moro de Cruzeiro, 35400-000 Ouro Preto, MG-Brasil
\\
$^{2}$Department of Applied Mathematics and Theoretical  Physics,
Cambridge University, Cambridge, UK
\\
$^{3}$NIKHEF, Amsterdam (Netherlands)
\\
$^{4}$ Leiden University, Leiden (Netherlands)
\\
$^{5}$Institute of Modern Physics, Chinese Academy of Sciences,
Lanzhou (China)
\\
$^{6}${\it Laboratoire de Math\'ematiques et de Physique Th\'eorique}, Tours University
(France)
\\
$^{7}$Faculty of Physics and Applied Informatics, University of Lodz, (Poland)
\\
$^{8}$
State Key Laboratory of Theoretical Physics, Institute of Theoretical Physics, Chinese Academy of Sciences, Beijing 100190, China 
}
\date{\today}

\begin{abstract}
The systematic derivation of constants of the motion, based on Killing tensors and the gauge covariant approach, is outlined. Quantum dots are shown to support second-, third- and fourth-rank Killing tensors. 
\\
%
%
\end{abstract}


\maketitle

\tableofcontents

\section{Introduction}

Noether's theorem associates a conserved quantity to a symmetry, defined as a transformation of space-time which changes the Lagrangian by a total derivative \ct{noether1918}. Infinitesimally, such symmetries are \emph{Killing vectors}. Higher-order expressions like the celebrated (Laplace-) Runge-Lenz vector cannot be obtained in this way, though, and require rather higher-rank generalizations called \emph{Killing tensors} \cite{Carter, 
KalninsMiller1980,KalninsMiller1981,Benenti1997,Benenti2005,DuvalValent}. The relations between these concepts become
particularly clear in the canonical formulation of the dynamics \cite{gibbons-rietdijk-vholten1993,jwvholten2006}. 
 
The (Laplace-) Runge-Lenz vector is associated with rank-2 tensors. Killing
tensors of rank $r \ge 2$ can also be considered, but physical examples are less common
\cite{Baleanu:1999zz,Igata,Gibbons:2011hg,Rugina,Galajinsky:2012vn,Visinescu}.  
 
The main result of this paper is a systematic derivation of higher-rank conserved quantities based on
Killing tensors, as illustrated by a \emph{fourth-rank} conserved quantity, \eqref{quarticCC} below. It is obtained for Quantum Dots for a particular choice of the parameters, when the system is integrable but not separable \cite{Simon,Alhassid,Blumel,ZZHG}.  
 Deriving this expression is far from being trivial : 
Bl\"{u}mel et al. \ct{Blumel}, e.g., found the correct formula at their second attempt only (and gave no
detailed explanation). The difficulty comes from that such higher-order expressions are, as
said above, not associated with a simple geometric action on space-time and therefore cannot be derived by the original Noether theorem. 
 
\section{Canonical approach to conservation laws}

For particles coupled to scalar and vector potentials $(\Fg, A_i)$ the hamiltonian takes the  form
\be
H = \frac{1}{2}\, g^{ij}(q)\, \Pi_i \Pi_j + \Fg(q), 
\label{2.1}
\ee
where the $\Pi_i=p_i-eA_i$ denote the gauge-covariant momenta; $e$ is the charge. The covariant
brackets read \ct{gibbons-rietdijk-vholten1993, jwvholten2006}
\be 
\left\{Q, K \right\} = D_i Q \dd{K}{\Pi_i} - \dd{Q}{\Pi_i} D_i K + e F_{ij}(q) \dd{Q}{\Pi_i} \dd{K}{\Pi_j},
\label{2.3}
\ee
where $F_{ij} = \der_i A_j - \der_j A_i$ is the field-strength tensor of $A_i$ on the configuration space 
with metric $g_{ij}$, and the covariant derivatives are defined by
\be
D_i K \equiv \ld \dd{K}{q^i} \right|_{\Pi} + \Gam_{ij}^{\;\;\;k}\, \Pi_k\, \dd{K}{\Pi_j}.
\label{2.4}
\ee
The definition reproduces the canonical equations of motion:
\beqa
\frac{dq^i}{dt} = \left\{ q^i, H \right\} = \dd{H}{\Pi_i} \hs{1} \Leftrightarrow \hs{1} 
\Pi_i = g_{ij} \frac{dq^j}{dt},
\label{2.5}
\\
\ba{l}
\dsp{ \frac{d\Pi_i}{dt} = \left\{ \Pi_i, H \right\} = - D_i H + e F_{ij} \dd{H}{\Pi_j} }
 \\
\dsp{ \hs{1} \Leftrightarrow \hs{1}
\frac{D\Pi_i}{Dt} \equiv \frac{d\Pi_i}{dt} - \frac{dq^j}{dt}\, \Gam_{ji}^{\;\;\;k}\, \Pi_k = e F_i^{\;j} \Pi_j - \dd{\Fg}{q^i}. }
\ea
\label{2.6}
\eeqa

Constants of the motion which are polynomials in the covariant momenta are  obtained by solving eq.
\beq
\left\{ Q, H \right\} = 0
\quad\hbox{
with}
\quad
Q = \sum_{n \geq 0} \frac{1}{n!} C^{(n)\, i_1 .... i_n}(q) \Pi_{i_1} ... \Pi_{i_n},
\label{3.0}
\ee
which leads to the generalized Killing equations
\be
\ba{l}
C^{(1)\, i}\, \Fg_{;i} = 0, \hs{2} C^{(0)}_{;i} = e\, C^{(1)\, j} F_{ij} + C^{(2)\, j}_{\;i} \Fg_{;j}, 
 \\[8pt]
C^{(n)}_{(i_1 ..i_n; i_{n+1})} = e\, C^{(n+1)\, j}_{(i_1...i_n} F_{i_{n+1})j} + C^{(n+2)\; j}_{i_1...i_{n+1}} \Fg_{;j}, 
  \hs{3} n \geq 1,
\ea
\label{3.1}
\ee
as discussed in Ref.\ \ct{jwvholten2006}. Note that any $Q$ which is a polynomial in $\Pi_i$ of rank $n$, 
has $C^{(m)} = 0$ for all $m > n$. In that case the highest coefficient tensor $C^{(n)}$ is a again a rank-$n$ 
Killing tensor:
\be
C^{(n)}_{(i_1 ... i_n; i_{n+1})} = 0; 
\label{3.2}
\ee
the next-to-highest coefficient tensor satisfies
\be
C^{(n-1)}_{(i_1 ... i_{n-1};i_n)} = e\, C^{(n)\, j}_{(i_1... i_{n-1}} F_{i_n) j}, 
\label{3.3}
\ee
and all tensors of rank $n-2$ and lower are subject to the full equation (\ref{3.1}).
Furthermore, the generalized Killing vector $C^{(1)}$ is always required to be orthogonal to 
the gradient of the scalar potential.

\section{Application to Quantum Dots}

The above formalism can be applied to the Quantum-Dot model of ref.\ \ct{Simon,Alhassid,Blumel,ZZHG}.
This concerns two charged particles with  Coulomb interaction in a constant magnetic field and a confining oscillator 
potential. The hamiltonian is 
\be
H = \sum_{a=1}^2 \left[ \frac{1}{2}\, \mathbf{\Pi}_a^2 + U(\mathbf{r}_a) \right] - \frac{a}{|\mathbf{r}_1 - \mathbf{r}_2|}.
\label{4.1}
\ee
The magnetic field direction is the $z$-direction, and the confining oscillator potential is taken to be axially symmetric: 
\be
U(\mathbf{r}_a) = \frac{1}{2} \left[ \og_0^2\, (x_a^2 + y_a^2) + \og_z^2\, z_a^2 \right]. 
\label{4.2}
\ee
Transformation to center-of-mass co-ordinates
$ 
\dsp{\mathbf{r}_{1,2} = \smallover{1}/{\sqrt{2}} \lh \mathbf{R} \pm \mathbf{r} \rh,}
$\;
$\dsp{\mathbf{\Pi}_{1,2} = \smallover{1}/{\sqrt{2}} \lh \, \mathbf{\Pi} \pm \bfpi \rh, }
$
leads to separation of variables,
$ 
H(\mathbf{r}_a, \mathbf{\Pi}i_a) = H_{CM}(\mathbf{R}, \mathbf{\Pi}) + H_{red}(\mathbf{r}, \bfpi),
$ 
with $H_{CM} = \frac{1}{2} \mathbf{\Pi}^2 + U(\bfR)$ and
\be
H_{red} = \frac{1}{2} \bfpi^2 + U(\mathbf{r}) - \frac{a}{r}.
\label{4.4}
\ee
As the magnetic field is constant,
$
F_{xy}=-F_{yx} = B, \, F_{yz} = F_{zx} = 0,
$
the covariant brackets separate as well:
$
\left\{ Q, K \right\} = \left\{Q, K \right\}_{CM} + \left\{ Q, K \right\}_{red}.
$
 In euclidean co-ordinates they read 
\be
\ba{l}
\dsp{ \left\{ K, Q \right\}_{CM} = \dd{K}{R^i} \dd{Q}{\Pi_i} - \dd{K}{\Pi_i} \dd{Q}{R^i} 
 + e B \lh \dd{K}{\Pi_x} \dd{Q}{\Pi_y} - \dd{K}{\Pi_y} \dd{K}{\Pi_x} \rh, }\\
 \\
\dsp{ \left\{ K, Q \right\}_{red} = \dd{K}{r^i} \dd{Q}{\pi_i} - \dd{K}{\pi_i} \dd{Q}{r^i} 
 + e B \lh \dd{K}{\pi_x} \dd{Q}{\pi_y} - \dd{K}{\pi_y} \dd{K}{\pi_x} \rh. }
\ea
\label{4.6}
\ee

 In the following we restrict ourselves to the effective 1-particle problem defined by the 
reduced brackets and hamiltonian $H_{red}$. To apply the formalism of generalized Killing
equations we lump the oscillator and Coulomb potential into the single scalar potential 
\be
\Fg = \frac{1}{2} \left[ \og_0^2 (x^2 + y^2) + \og_z^2 z^2 \right] - \frac{a}{\sqrt{x^2 + y^2 + z^2}}.
\label{4.7}
\ee
The equations of motion in 3D euclidean co-ordinates derived from the reduced hamiltonian and brackets then read
\be
\dot{r}_i = \left\{ r_i, H \right\} = \pi_i, \hs{2}
\dot{\pi}_i = \left\{ \pi_i, H \right\} = - \Fg_{,i} + \eps_{ijz} e B \pi_j,
\label{4.8}
\ee
where the comma denotes an ordinary partial derivative w.r.t.\ $r_i$, and $\eps_{ijk}$ is the permutation tensor. 
The corresponding quantum theory is obtained by replacing the brackets of phase-space functions $(K, Q)$ by 
operator commutation relations.

To make use of axial symmetry, it is convenient to transform to curvilinear
co-ordinates $\xi^i = (\rg, z, \vf)$. Then the hamiltonian becomes
\be
H = \frac{1}{2} g^{ij} \pi_i \pi_j + \Fg,
\label{5.1}
\ee
with metric $g_{ij} =$ diag$(1,1, \rg^2)$ and scalar potential (\ref{4.7}). 
As $F_{\rg \vf} = \rg B$, the covariant brackets are given by
\be
\left\{ K, Q \right\} = D_i K\, \dd{Q}{\pi_i} - \dd{K}{\pi_i}\, D_i Q 
  + e B \rg \lh \dd{K}{\pi_{\rg}} \dd{Q}{\pi_{\vf}} - \dd{K}{\pi_{\vf}} \dd{Q}{\pi_{\rg}} \rh,
\label{5.3}
\ee
where the covariant derivatives $D_i$ are defined by (\ref{2.4}) with non-zero connection coefficients
$ 
\Gam_{\rg\vf}^{\;\;\;\vf} = {1}/{\rg}, \,
 \Gam_{\vf\vf}^{\;\;\;\rg} = - \rg.
$
As the angle $\vf$ is a cyclic co-ordinate, the corresponding canonical momentum is conserved. 
In the present covariant formalism this follows from the existence of a Killing vector and a Killing scalar,
$ 
\lh C^{(1)\, \rg}, C^{(1)\, z}, C^{(1)\, \vf} \rh = (0, 0, 1),\,
C^{(0)} = \og_L \rg^2$, respectively, where $\og_L = {eB}/{2}
$.
They combine into the constant of the motion, namely the $z$-component of the total angular momentum,
\be
L_z = C^{(1)\, i} \pi_i + C^{(0)} = \pi_{\vf} + \og_L \rg^2.
\label{5.7}
\ee
The Hamilton equations 
$
{d\xi^i}/{dt} = \left\{\xi^i, H \right\} = g^{ij} \pi_j 
$
imply
$ 
{d\vf}/{dt} = {\pi_{\vf}}/{\rg^2}
 = {L_z}/{\rg^2} - \og_L;
$
 $\og_L$ is hence the Larmor frequency. 

As discussed in \ct{Simon,Alhassid,Blumel,ZZHG}, the model allows for more constants of 
motion whenever certain specific conditions on the frequencies
 hold, namely for certain exceptional values  of $\tau=\omega_z/\sqrt{\omega_0^2+\omega_L^2}\,$. 
 
$\bullet$ We first observe that there is a \emph{rank-2 Killing tensor}
\be
C^{(2)}_{ij} = \lh \ba{ccc} 2z & -\rg & 0 \\
- \rg & 0 & 0 \\ 
 0 & 0 & 2 \rg^2 z \ea \rh   \hs{1} \Leftrightarrow \hs{1} 
\frac{1}{2} C^{(2)\, ij} \pi_i \pi_j = z \pi^2_{\rg} - \rg \pi_{\rg} \pi_z + \frac{z}{\rg^2} \pi^2_{\vf}.
\label{5.10}
\ee
Then the generalized Killing equation (\ref{3.3}) for $C^{(1)}$ is solved by
\be
C^{(1)}_i = \lh 0, 0, eB\rg^2 z \rh \hs{1} \Leftrightarrow \hs{1} C^{(1)\,i} \pi_i = e B z \pi_{\vf}.
\label{5.11}
\ee
This vector is orthogonal to the gradient of $\Fg$, as required by the first equation in (\ref{3.1}).
Finally, the Killing scalar $C^{(0)}$ must satisfy
\be
\ba{l}
\dsp{ C^{(0)}_{, \rg} = \lh 4 \og_L^2 + 2 \og_0^2 - \og_z^2 \rh \rg z + \frac{a \rg z}{(\rg^2 + z^2)^{3/2}}, }
\\[6pt]
\dsp{ C^{(0)}_{, z} = - \og^2_{\rg} \rg^2 - \frac{a \rg^2}{(\rg^2 + z^2)^{3/2}}, 
\hs{2} C^{(0)}_{,\vf} = 0. }\\
\ea
\label{5.12}
\ee
A solution, namely
\be
C^{(0)} = - \og^2_{\rg} \rg^2 z - \frac{az}{\sqrt{\rg^2 + z^2}}\,, 
\label{5.13}
\ee
exists, provided the frequencies satisfy
\be
\og_{0}^2 + \og_L^2 = \frac{1}{4} \og_z^2.
\label{5.14}
\ee
Combining the results, the full constant of motion,
\be
Q_1 = z \pi_{\rg}^2 - \rg \pi_{\rg} \pi_z + \frac{z}{\rg^2} \pi_{\vf}^2 + 2 \og_L z \pi_{\vf}
 - \og_0^2 \rg^2 z - \frac{az}{\sqrt{\rg^2 +z^2}}, 
\label{5.15}
\ee
We recover hence the  Runge-Lenz-type quadratic expression found before
for $\tau=2$, when the system is  separable in parabolic coordinates \cite{Simon,Alhassid,Blumel,ZZHG}.

Another constant can be constructed starting from the \emph{rank-4 Killing tensor}
\be
\frac{1}{4!} C^{(4) ijkl} \pi_i \pi_j \pi_k \pi_l = 
 \rg^2 \pi_z^4 - 2 \rg z \pi_{\rg} \pi_z^3 + z^2 \pi_{\rg}^2 \pi_z^2
  + \frac{1}{\rg^2}\, \pi_{\vf}^4 + \pi_{\rg}^2 \pi_{\vf}^2 + \lh 2 + \frac{z^2}{\rg^2} \rh \pi_z^2 \pi_{\vf}^2.
\label{5.16}
\ee
Then solving eqs.\ (\ref{3.3}) for $C^{(3)}$ one finds the minimal solution
\be
\frac{1}{3!} C^{(3)\,ijk} \pi_i \pi_j \pi_k = 2 \og_L \pi_{\vf} \lh \rg^2 \pi^2_{\rg} + (2 \rg^2 + z^2) \pi^2_z \rh, 
\label{5.17}
\ee
modulo a rather long list of separate rank-3 Killing tensor expressions, defining independent constants of motion. These are discussed in  eq. (\ref{6.1}) below.

The terms quadratic in the covariant momenta are now obtained in a straightforward way by requiring all
contributions of order $\pi^3$ to the bracket $\left\{ Q, H \right\}$  to cancel; this gives the minimal expression,
\be
\ba{lll}
\dsp{ \frac{1}{2}\, C^{(2)\, ij} \pi_i \pi_j }& = & 
 \dsp{ \left[ (2 \og_z^2 - \og_0^2) z^2 \rg^2 + 2 \og_L^2\rg^4 - \frac{2a \rg^2}{\sqrt{z^2 + \rg^2}} \right] \pi_{z}^2 }\\[12pt]
 & & \dsp{ +  \left[ \frac{2}{3} \lh 2 \og_0^2 - 5 \og_z^2 + 2 \og_L^2 \rh z^3 \rg + \frac{2az\rg}{\sqrt{z^2 + \rg^2}} \right] 
  \pi_z \pi_{\rg} }
\\[12pt]
 & & \dsp{ + \left[ \og_L^2 \rg^4 - \frac{1}{3} \lh \og_0^2 - 4 \og_z^2 + \og_L^2 \rh z^4 \right] \pi_{\rg}^2 }
\\[12pt]
 & & \dsp{ + \left[ 2 \og_z^2 z^2 + \lh \og_0^2 - 5 \og_L^2 \rh \rg^2 - \frac{1}{3} \lh \og_0^2 - 4 \og_z^2 + \og_L^2 \rh 
  \frac{z^4}{\rg^2} - \frac{2a}{\sqrt{z^2 + \rg^2}} \right] \pi_{\vf}^2. }              
\ea
\label{5.18}
\ee
Calculating the contributions of order $\pi^2$, we have to add a linear term
\be
\ba{lll}
C^{(1) i} \pi_i \hs{-.5} & = & \dsp{ \hs{-.5} -2 \og_L \pi_{\vf} \left[ \frac{1}{3} \lh \og_0^2 - 4 \og_z^2 + \og_L^2 \rh z^4 
 - 2 \og_z^2 z^2 \rg^2 + \lh 3 \og_L^2 - \og_0^2 \rh \rg^4 + \frac{2a \rg^2}{\sqrt{z^2 + \rg^2}}  \right]. } 
\ea 
\label{5.19}
\ee
It remains to find a $C^{(0)}(z,\rg)$ such that the bracket closes. This requires
\be
\ba{lll}
C^{(0)}_{\; ,z} & = & \dsp{ \lh 4 \og_z^4 - \frac{16}{3} \og_z^2 \og_\rg^2 + \frac{4}{3} \og_0^4 + 
 \frac{4}{3} \og_L^2 \og_0^2 \rh z^3 \rg^2 + 4 \og_L^2 \og_z^2  z \rg^4 - \frac{2a^2 z \rg^2}{(z^2 + \rg^2)^2} }
 \\[8pt]
 & & \dsp{ +\; \frac{a}{(z^2 + \rg^2)^{3/2}} \left[ \lh - \frac{10}{3} \og_z^2 + \frac{4}{3} \og_0^2 + \frac{4}{3} \og_L^2 \rh z^3 \rg^2
+ \lh - 4 \og_z^2 + 2 \og_0^2 + 4 \og_L^2 \rh z \rg^4 \right], }\\
 & & \\
C^{(0)}_{\; ,\rg} & = & \dsp{ \lh - \frac{10}{3} \og_z^4 + 4 \og_z^2 \og_0^2 - \frac{2}{3} \og_0^4 + 
 \frac{20}{3} \og_L^2 \og_z^2 - 2\og_L^2 \og_0^2 - \frac{4}{3} \og_L^4 \rh z^4 \rg }
 \\[8pt]
 & & \dsp{ +\; 8 \og_L^2 \og_z^2 z^2 \rg^3 - 6 \og_L^2\lh 2 \og_L^2 - \og_0^2 \rh \rg^5 + \frac{2a^2 z^2 \rg}{(z^2 + \rg^2)^2} }
 \\
 & & \dsp{ +\; \frac{a}{(z^2 + \rg^2)^{3/2}} \left[ \frac{2}{3} \lh \og_0^2 + 2 \og_z^2 + \og_L^2 \rh z^4 \rg
  + 2 \lh \og_z^2 - 4 \og_L^2 \rh z^2 \rg^3 - 6 \og_L^2 \rg^5 \right]. }
\ea
\label{5.20}
\ee
Let us first turn off the Coulomb potential, i.e., consider the $a$-independent part of these eqns. Then the integrability condition is 
\be
\lh 2 \og_L^2 -5 \og_z^2 + 2 \og_0^2 \rh^2 = 9 \og_z^4, 
\label{5.21}
\ee
with allows for two solutions, namely
\be
\mbox{(a)} \hs{1} \og_L^2 = 4 \og_z^2 - \og_0^2, \hs{2} 
\mbox{(b)} \hs{1} \og_L^2 = \og_z^2 - \og_0^2.
\label{5.22}
\ee
Thus, for the \emph{ magnetic problem with a confining harmonic potential but with no Coulomb potential}, there are two values of the Larmor frequency $\og_L$ for which there
is a quartic constant of motion. 

$\bullet$ In contrast, when $a \neq 0$, i.e., when the \emph{Coulomb potential is switched on}, the terms linear in $a$ are integrable only if condition (a) 
is satisfied (terms proportional to $a^2$ are actually always integrable)
 -- which is, indeed, the integrable but non-separable case $\tau=1/2$ in \cite{ZZHG} cf. \cite{Simon,Alhassid,Blumel,ZZHG}. 
Imposing condition (a), the minimal solution for 
$C^{(0)}$ becomes
\be
\ba{lll}
C^{(0)} & = & \dsp{ \og_z^4 z^4 \rg^2 + 2 \og_z^2 \og_L^2 z^2 \rg^4 - \og_L^2 \lh 3 \og_L^2 - 4 \og_z^2 \rh \rg^6 }
\\[4pt]
 & & \dsp{ + \frac{2a}{\sqrt{z^2 + \rg^2}} \left[ \og_z^2 z^2 \rg^2 - \og_L^2 \rg^4 \right] 
- \frac{a^2}{2} \frac{z^2 - \rg^2}{z^2 + \rg^2},}
\ea
\label{5.23}
\ee
with $\og_L^2 + \og_0^2 = 4 \og_z^2$. Then the sum of the expressions (\ref{5.16}), (\ref{5.17}), (\ref{5.18}), (\ref{5.19})
and (\ref{5.23}) represents, for this special value of the magnetic
field, the quartic constant of the motion in the CM system. Explicitly,
\beq
\ba{lll}
Q_2 & = & \dsp{ \rg^2 \pi_z^4 - 2 \rg z \pi_{\rg} \pi_z^3 + z^2 \pi_{\rg}^2 \pi_z^2
  + \frac{1}{\rg^2}\, \pi_{\vf}^4 + \pi_{\rg}^2 \pi_{\vf}^2 + \lh 2 + \frac{z^2}{\rg^2} \rh \pi_z^2 \pi_{\vf}^2 }
\\[6pt]
 & & \dsp{ +\, 2 \og_L \pi_{\vf} \lh \rg^2 \pi^2_{\rg} + (2 \rg^2 + z^2) \pi^2_z \rh + 
   \left[ (2 \og_z^2 - \og_0^2) z^2 \rg^2 + 2 \og_L^2\rg^4 - \frac{2a \rg^2}{\sqrt{z^2 + \rg^2}} \right] \pi_{z}^2 }
\\[6pt]
 & & \dsp{ +  \left[ \frac{2}{3} \lh 2 \og_0^2 - 5 \og_z^2 + 2 \og_L^2 \rh z^3 \rg + \frac{2az\rg}{\sqrt{z^2 + \rg^2}} \right] 
  \pi_z \pi_{\rg} + \left[ \og_L^2 \rg^4 - \frac{1}{3} \lh \og_0^2 - 4 \og_z^2 + \og_L^2 \rh z^4 \right] \pi_{\rg}^2 }
\\[12pt]
 & & \dsp{ + \left[ 2 \og_z^2 z^2 + \lh \og_0^2 - 5 \og_L^2 \rh \rg^2 - \frac{1}{3} \lh \og_0^2 - 4 \og_z^2 + \og_L^2 \rh 
  \frac{z^4}{\rg^2} - \frac{2a}{\sqrt{z^2 + \rg^2}} \right] \pi_{\vf}^2 }
\\[8pt]
 & & \dsp{ -\, 2 \og_L \pi_{\vf} \left[ \frac{1}{3} \lh \og_0^2 - 4 \og_z^2 + \og_L^2 \rh z^4 
 - 2 \og_z^2 z^2 \rg^2 + \lh 3 \og_L^2 - \og_0^2 \rh \rg^4 + \frac{2a \rg^2}{\sqrt{z^2 + \rg^2}}\right] }
\\[8pt]
  & & \dsp{ +\, \og_z^4 z^4 \rg^2 + 2 \og_z^2 \og_L^2 z^2 \rg^4 - \og_L^2 \lh 3 \og_L^2 - 4 \og_z^2 \rh \rg^6 }
 \dsp{  +\, \frac{2a}{\sqrt{z^2 + \rg^2}} \left[ \og_z^2 z^2 \rg^2 - \og_L^2 \rg^4 \right] 
  - \frac{a^2}{2} \frac{z^2 - \rg^2}{z^2 + \rg^2}\,,}
\ea 
\label{quarticCC}
\eeq 
which is in fact the conserved quantity found in the integrable-but-non-separable case 
$\tau=1/2$ using quite different methods \cite{ZZHG}.
By construction, the coefficients of the quartic term (\ref{5.16}) define a rank-4 Killing tensor w.r.t.\ the metric (\ref{5.1}). 

More generally, the complete list of rank-3 Killing tensors is
\be
\ba{l}
K_1^{(3)} = \pi_z^3, \hs{8} K_2^{(3)} = \pi_{\vf}^3, \\
K_3^{(3)} = \pi_z \pi_{\vf}^2, \hs{7}  \dsp{ K_4^{(3)} = \pi_z \lh \pi_{\rg}^2 + \frac{1}{\rg^2} \pi_{\vf}^2 \rh, }
\\[12pt]
\dsp{ K_5^{(3)} = \pi_{\vf} \lh \pi_{\rg}^2 + \frac{1}{\rg^2} \pi_{\vf}^2 \rh,} \hs{2} 
\dsp{ K_6^{(3)} = \pi_z \left[ - \rg \pi_{\rg} \pi_z  + z \lh \pi_{\rg}^2 + \frac{1}{\rg^2} \pi_{\vf}^2 \rh \right], }
\\[12pt]
\dsp{K_7^{(3)} = \pi_z \left[ \frac{1}{2} \rg^2 \pi_z^2 - z \rg \pi_{\rg} \pi_z + \frac{1}{2} z^2 \lh \pi_{\rg}^2
 + \frac{1}{\rg^2} \pi_{\vf}^2 \rh \right]. }
\ea
\label{6.1}
\ee
Note, however, that these are composed of direct products of lower-rank Killing tensors and vectors. 
In particular, $K^{(1)}_z = \pi_z$ and $K^{(1)}_{\vf} = \pi_{\vf}$ define Killing vectors by themselves:
$
K^{(1)\, i}_z = (0, 1, 0), \, K^{(1)\, i}_{\vf} = (0, 0, 1).
$ 

  The expressions in eqs.\ (\ref{6.1}) are products of these Killing vectors with 
each other and with the rank-2 Killing tensors
\be
\ba{l}
\dsp{ K^{(2)}_1 = \pi_{\rg}^2 + \frac{1}{\rg^2} \pi_{\vf}^2, \hs{2} 
K^{(2)}_2 = - \rg \pi_{\rg} \pi_z  + z \lh \pi_{\rg}^2 + \frac{1}{\rg^2} \pi_{\vf}^2 \rh },
 \\[4pt]
\dsp{ K^{(2)}_3 =  \frac{1}{2} \rg^2 \pi_z^2 - z \rg \pi_{\rg} \pi_z + \frac{1}{2} z^2 \lh \pi_{\rg}^2
 + \frac{1}{\rg^2} \pi_{\vf}^2 \rh. }
\ea
\label{6.3}
\ee
We discuss these expressions in turn. First, $K_1^{(2)}$ satisfies the bracket relation
\be
\left\{ K^{(2)}_1, H \right\} = - 2 \rg \pi_{\rg} \lh \og_0^2 + \frac{a}{(z^2 + \rg^2)^{3/2}} \rh.
\label{6.4}
\ee
Now as $\pi_{\rg}$ is not a Killing vector, $K_1^{(2)}$ can be turned into a constant of motion only 
by adding a scalar term $K_1^{(0)}$ such that 
\be
K^{(0)}_{1\, ,z} = 0, \hs{2} K^{(0)}_{1\, ,\rg} = 2 \rg \lh \og_0^2 + \frac{a}{(z^2 + \rg^2)^{3/2}} \rh.
\label{6.5}
\ee
The solution of the 2nd equation:
\be
K_1^{(0)} = \og_0^2 \rg^2 - \frac{a}{\sqrt{z^2 + \rg^2}},
\label{6.6}
\ee
satisfies the first equation (\ref{6.5}) only if $a = 0$. Therefore this quadratic Killing tensor generates a 
constant of motion only in the absence of a Coulomb potential: $a = 0$.

An alternative is to replace the 
3D Coulomb potential by a 2D one:
\be
\Fg \rightarrow \tilde{\Fg} = \frac{1}{2} \og_z^2 z^2 + \frac{1}{2} \og_0^2 \rg^2 - \frac{a}{\rg}.
\label{6.7}
\ee
Next, observe that $K^{(2)}_2$ is identical to the Killing tensor (\ref{5.10}). We have already seen that
it can be extended to a constant of motion only if the Larmor frequency is tuned to take the value (\ref{5.14}). 

Finally, we  discuss $K_3^{(2)}$.  A straightforward calculation along the previous lines shows that
it can also be extended to a complete constant of motion, 
\be
Q_3 = \frac{1}{2} \rg^2 \pi_z^2 - z \rg \pi_z \pi_{\rg}  + \frac{1}{2} z^2 \lh \pi_{\rg}^2 + \frac{1}{\rg^2} \pi_{\vf}^2 \rh
  +\, \og_L z^2 \pi_{\vf} + \frac{1}{2} \og_L^2 z^2 \rg^2, 
\label{6.8}
\ee
provided 
$
\og_L^2 = \og_z^2 - \og_0^2,
$
--- which is condition (b) in (\ref{5.22}).

We observe that (\ref{6.8}) 
 is in fact the difference of two separately conserved quantities found in Ref. \cite{ZZHG}, $\half(L^2-L_z^2)$,
i.e., (half of) the \emph{total angular momentum squared}, $L^2$, minus $L_z^2$, the \emph{square of the third component of} $L_z$. 
This is no surprise since condition (b) in (\ref{5.22}) means $\tau=1$, which amounts to \emph{spherical symmetry} and hence \emph{conserved total angular momentum} after elimination of the magnetic field \ct{ZZHG}.

\goodbreak\newpage
\section{Algebraic Structure\label{sec:algebraic_structure}} 
 
When we have several symmetries, their algebraic structure is of fundamental importance. (Remember the $\orth(4)/\orth(3, 1)$ dynamical symmetry of the Kepler problem.) 
How do the Killing tensors reflect the Poisson algebra structure of the associated conserved quantities?

The answer is non-trivial let alone in the simplest, rank-1 case, when the  Poisson structure of the conserved quantities may not be the same
as that of the generating vectors under Lie bracket. Just consider a free particle: while the vectors of the infinitesimal action on space-time span the center-less Galilei Lie algebra, the associated conserved quantities realize its central extension (called the Bargmann algebra).

 This problem can be conveniently dealt with using a higher-dimensional framework \cite{DGH91} and Schouten-Nijenhuis algebras  \ct{Gibbons:2011hg}. Full details will be presented elsewhere. Here we satisfy ourselves with some general remarks about the bracket algebra \ct{jwvholten1997}. Let $J^{(p)}$ be constructed
from a highest-rank Killing tensor of rank p. By construction, the bracket of two
such constants of motion has the general structure  
\be 
\{ J^{(p)}, J^{(q)} \} \sim J^{(p+q-1)} \, . 
\ee 
It follows that the generators of rank $p = 1$ form a closed algebra, namely a Lie algebra if the
structure functions are constant. The constants of rank $p \ge 1$ then must form representations of this algebra: 
\be 
\{ J^{(p)} , J^{(1)} \} \sim J^{(p)} \, . 
\ee 
If there is more than one constant of the motion of rank $p, q \ge 2$, their bracket generates
constants of motion of higher rank $p + q - 1 \ge p + 1$. Therefore either the $J^{(p)}$, $p \ge 2$,
form a trivial representation of the Lie algebra and all their brackets vanish, or an infinite-dimensional
set of constants of the motion is generated. Well-known examples of such
infinite-dimensional algebras are the Virasoro and Kac-Moody algebras. However, these
non-trivial infinite-dimensional algebras arise only for infinite-dimensional systems. In the
finite-dimensional case we expect the brackets of higher-rank constants of motion to vanish,
or represent simple powers and products of lower-rank constants. 
 
Obviously, the Lie-algebra of constants $J^{(1)}$ generates transformations in configuration
space, and corresponding configuration-dependent linear transformations in momentum
space separately: 
\be 
J^{(1)} = \xi^i (q) p_i \quad \Rightarrow \quad \{ q^i , J^{(i)} \} = \xi^i (q) \, , \quad  \{ q^i , J^{(i)} \}  = - \partial_i \xi^j (q) p_j \, . 
\ee 
In contrast, the higher-rank constants generate transformations in phase space, which mix
configuration- and momentum variables $(q^i, p^i)$ -- just like in the Kepler case.

\section{Conclusion}

To summarize our results, we derived, after outlining a general covariant framework based on 
 higher-rank Killing tensors, three constants of the motion for the Quantum Dot Hamiltonian (\ref{5.1}) with trapping potential (\ref{4.7}). These additional conserved quantities  arise for specific values of the magnetic field strength~$B$.  Note that terms which are odd in the momenta are multiplied by odd powers of the Larmor frequency consistently with \emph{time-reversal symmetry}.

 Case (\ref{5.14}) i.e. $\og_L^2 = \frac{1}{4}\og_z^2-\og_0^2$ corresponds to $\tau=2$ which is the one separable in parabolic coordinates and admits the \emph{quadratic} ``Runge-Lenz-type scalar'' constant of motion (\ref{5.15}).

For Case (a) in (\ref{5.22}) i.e., for
$\og_L^2 = 4 \og_z^2 - \og_0^2$,
there is a \emph{quartic} constant of the motion, namely $Q_2$ (\ref{quarticCC}), consistent with the expression found in Ref. \ct{ZZHG} in the integrable but non-separable case
$\tau=1/2$.

In Case (b) in (\ref{5.22}), i.e., for
$\og_L^2 = \og_z^2 - \og_0^2,$
we have another quadratic constant of motion, namely $Q_3$ in (\ref{6.8}),
which is in fact the difference of the total angular momentum squared minus the square of the third component, found in Ref. \cite{ZZHG} for $\tau=1$ i.e., when the system is hiddenly and effectively spherically symmetric. 
  
In the  the integrable cases $\tau = 1, 2, 1/2$ we have found a maximal set of mutually commuting conserved quantities. No other independent and mutually commuting quantities can be found. However, this does not exclude \textit{a priori} the possibility that the system is super-integrable and that further conserved quantities exist, that would not commute with the ones we had found. Our systematic analysis implies that any such further conserved quantity, if it does exist, must be either of order $>4$ or non-polynomial in the momenta. 

At last, one might wonder if one could not start with a systematic determination of all Killing tensors. This should in principle be possible by solving the Killing equations for any given rank, generalizing the standard procedure followed for Killing vectors. This is a rather laborious task, though, and the only results we are aware of concern rank-two tensors in the free case, with a huge number of Killing tensors \cite{Durand}. 
  
More details and examples are discussed in a follow-up paper \ct{CGvHHZ2014}.

\begin{acknowledgments}
For JWvH this work is part of the research program of the Foundation for Research of Matter (FOM).
P.K. would like to acknowledge  Pawel Maslanka and Cezary Gonera for discussions.
This work was partially supported also by the National Natural Science Foundation of
China (Grants No. 11035006 and 11175215) and by the Chinese Academy of Sciences Visiting
Professorship for Senior International Scientists (Grant No. 2010TIJ06).

\end{acknowledgments}


\end{document}